\documentclass[twocolumn,aps,prb,showpacs,superscriptaddress,longbibliography]{revtex4-1}

\usepackage{graphicx}
\usepackage{dcolumn}
\usepackage{bm}
\usepackage{color}
\usepackage{siunitx}
\usepackage{xcolor}

\usepackage{graphicx}
\usepackage{dcolumn}
\usepackage{bm}
\usepackage{color}

\pagecolor{white}

\begin{document}

\title{Dipolariton propagation in a Van der Waals TMDC with $\Psi$-shaped channel guides and buffered channel branches}

\author{Patrick Serafin}
\affiliation{Physics Department, New York City College of Technology, The City University of New York, Brooklyn, New York 11201, USA}

\author{German V. Kolmakov}
\affiliation{Physics Department, New York City College of Technology, The City University of New York, Brooklyn, New York 11201, USA}

\date{\today}

\begin{abstract}
Using a computational approach based on the driven diffusion equation for dipolariton wave packets, we simulate the diffusive dynamics of dipolaritons in an optical microcavity embedded with a transition-metal dichalcogenide (TMDC) heterogeneous bilayer encompassing a  $\Psi$-shaped channel.  By considering  exciton-dipolaritons, which are a three way superposition of direct excitons, indirect excitons and cavity photons, we are able to drive the dipolaritons in our system by the use of an electric voltage and investigate their diffusive properties. More precisely, we study the propagation of dipolaritons present in a MoSe$_2$-WS$_2$  heterostructure, where the dipolariton propagation is guided by a  $\Psi$-shaped channel. We also consider the propagation of dipolaritons in the presence of a buffer in the $\Psi$-shaped channel and study resulting changes in efficiency.  We introduce designs for optical routers at room temperature as well as show that system parameters including driving forces of $\approx$2.0eV/mm and electric field angles of sixty degrees optimize the dipolariton redistribution efficiencies in our channel guide.

\end{abstract}

\maketitle

\section{Introduction}

	Recent advances in the field of polaritronics have enabled us to realize many proposed application of exciton and polariton physics at room temperature scales \cite{tsintzos:5} Exciton-polariton physics has already found its application in devices such as lasers \cite{Christopoulos:10}, optical transistors  \cite{Serafin17,Gao:85} and light emitting diodes \cite{jayaprakash:7}. The wide array of phenomena found in exciton-polariton physics, such as superfluidity \cite{MacDonald:08,Berman:08,Berman:10,Balili} and Bose-Einstein condensate \cite{byrnes:91,Richard7} has drawn attention towards research into their implementation in various systems as well as potential applications \cite{Sanvitto:4}. There have been many experimental developments regarding exciton-polariton confinement\cite{Kim2} and trapping\cite{Balili} as well as the ability to make the condensate flow\cite{Su} paving the way towards further inquiry into the control and manipulation of exciton-polaritons.

	Some obstacles that can arise with the use of conventional polaritons, such as low exciton binding energies and electrical neutrality can be remedied by the use of charged quasi-particles; that is, dipolaritons in van der Waals transition metal dichalcogenide (TMDC) heterostructures. Dipolaritons present us with the possibility of being driven by an external voltage, something that is not possible in a conventional polariton as it is electrically neutral. The emergence of Van der Waal TMDC heterostructures enables us to use properties such as large binding energies and large Rabi splitting energies to transfer proposed applications of polaritons to room temperature scales. In addition, these TMDC heterostructures provide for a type-II band alignment \cite{Ceballos:7} which enables one to spatially separate electrons and their respective holes upon laser pumping into the system.

	In this paper, the diffusive dynamics of dipolaritons present in a TMDC inside of an optical microcavity is computationally modeled. In particular, we consider  heterostructures with a $\Psi$-shaped channel guide in contrast to heterostructures with Y-shaped channel guides previously investigated for identical system parameters \cite{Serafin17}. This MoSe$_2$-WS$_2$ heteregenous bilayer embedded inside of a optical microcavity enables one to redistribute dipolaritons through the $\Psi$-shaped channel guide using an external voltage as shown in Fig.1. The system parameters for this redistribution are investigated in order to determine the conditions that optimize the efficiency of redistribution of the dipolariton gas in our sytem. By considering a novel $\Psi$-shaped channel guide, we are able to compare efficiency between geometrically different channel guides. In particular, by considering various driving forces, electric field angles, and channel buffers we can establish optimal system parameters that promote efficient dipolariton routing in the $\Psi$-shaped channel of the TMDC.

 \begin{figure}[h]
\begin{center}
\includegraphics[height=8cm, width=13cm]{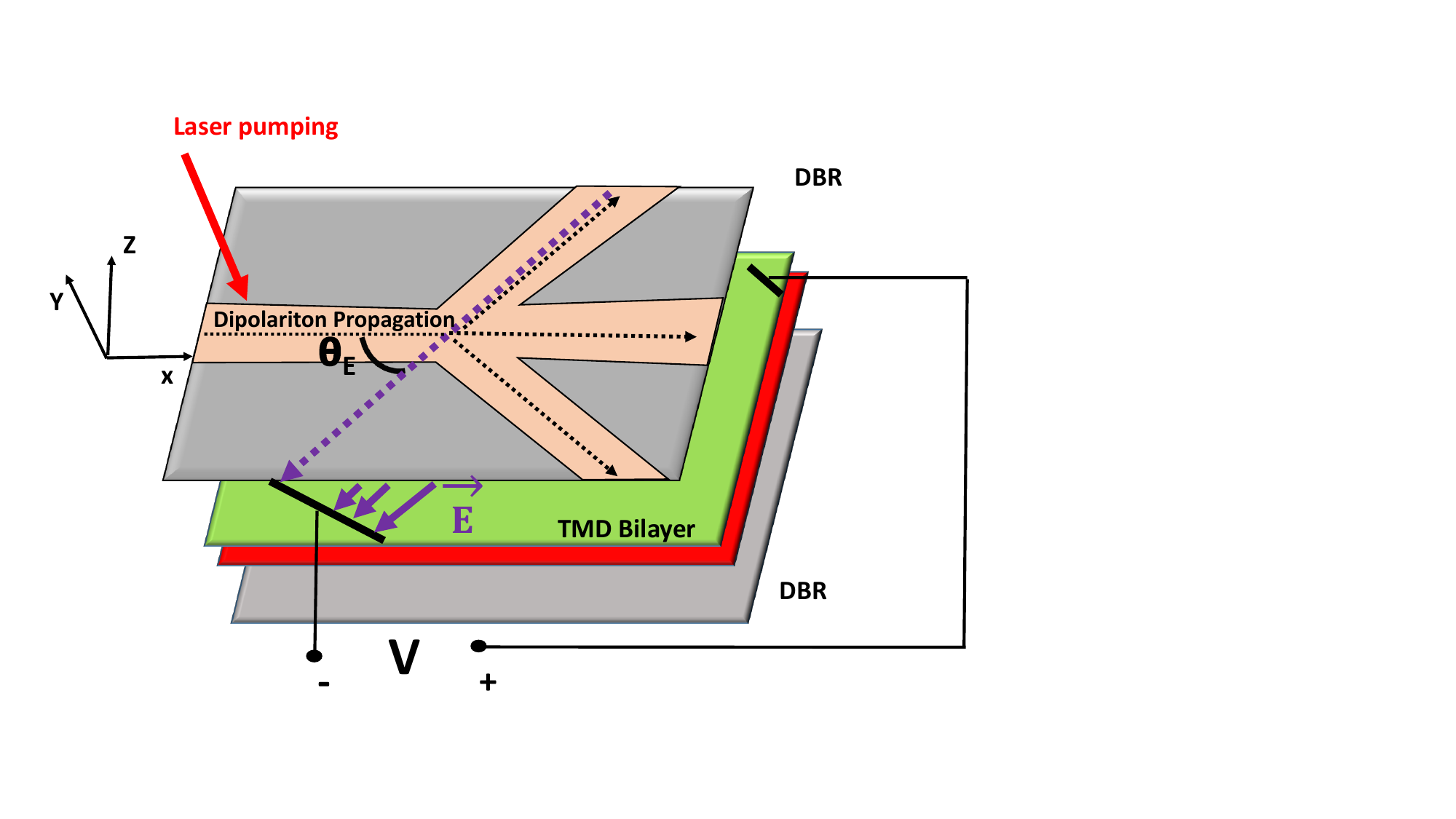}
\caption{Schematic of a TMD heterogenous bilayer embedded inside an optical microcavity with a $\Psi$-shaped channel guiding the dipolaritons. The opening angle of the channel relative to the middle branch is $\theta_0=30$\ and the direction of the field $\bm{E}$ generated by an external voltage applied to the bilayer is defined by an angle $\theta_E$ between the field vector and the direction of the stem of the  $\Psi$-shaped channel. The direction of the electric driving force applied to the dipolaritons is opposite to the direction of the electric field. Distributed Bragg Reflectors (DBR) are placed between the TMD bilayer and laser pumping is applied to generate dipolaritons, at the beginning of the stem of the channel at which point they propagate along the $x$-axis  towards the junction under the action of the electric field $\bm{E}$.}
\label{fig:pattern}
\end{center}
\end{figure}

\section{Simulation Method}
\label{sec:method}
 
	We model our system of dipolariton propagation in a TMDC with a $\Psi$-shaped channel using previously considered models for dipolariton based optical transistors using TMDC's \cite{Serafin17}. In particular, we consider the propagation of dipolaritons controlled by an external electric field $E$ generated by an external voltage acting on the charges found in the MoSe$_2$ layer of the TMDC heterostructure as seen in Fig.1. The polariton source is generated at the stem on the left side of the $\Psi$-shaped channel as seen in Fig. 1. whereby direct excitons are created in the excitation spot of characteristic size 10-20$\mu$m. More precisely in our model, polaritons are added at time intervals dt with a Gaussian spatial distribution at the laser spot center located at the base of the stem of the channel on the left side of Fig. 1. The voltage generated acts upon the charges in the MoSe2 layer of the TMDC, thus enabling us to control the routing of dipolaritons in the channel.  To describe this stochastic model for the dipolariton gas we use the Langevin equation \cite{Ohzeki_2015} for the center of mass coordinate \textbf{r}(t) of polariton wave packets\cite{Kolmakov:33}
	
\begin{equation}  
d\bm{r}= \eta_{\text{dip}} \bm{F}(\bm{r}(t),t))dt + \sqrt{2D}d\bm{\xi}(t)		
\end{equation}

 The first term in Eq.(1) $\bm{F}(\bm{r}(t))$ represents the external force acting on the dipolaritons \cite{Kim2}, the second term $\eta_{\rm dip}$  is the dipolariton mobility, whilst $D$ is the diffusivity, and $\bm{\xi}(t)$ the differential of a Wiener process \cite{Kolmakov:33}. The reason for modeling our system with Eq. (1), is that it models fluctuation of particle density in the polariton gas and has shown to provide computational effectiveness for a system of large particles and those possessing off lattice properties \cite{Deng199}. Eq.(1) has been applied in other models such as ones studying the diffusive properties of gases \cite{Carusotto2013} and nanoparticles in fluids \cite{Ladd10}. Eq. (1) does not follow a detailed balance condition in our case due to non-zero fluxes of particles and damping our state is not in equilibrium and thereby does not satisfy this condition\cite{Ohzeki_2015}. 
 
This computational model adhreres to the same physics and identical system parameters as described when considering this system with a Y-shaped channel guide \cite{Serafin17}. The model parameters are summarized in Table 1 and their properties and derivations will be briefly reviewed here.

 Our simulations for the dipolariton dynamics for our system are taken at room temperature; that is, T = 300 K. The lower dipolariton mass is calculated as follows
\begin{equation}
{1 \over m}= {|C|^2 \over m_{\rm ph}} + {|X|^2+|Y|^2 \over m_{\rm ex}}.
\end{equation}

In this equation, $m_{\rm ph}$ is the effective cavity photon mass, $m_{\rm ex}$=$m_{\rm e}$+$m_{\rm h}$ is the exciton mass 
$m_{\rm e(h)}$ is the electron (hole) mass,  where we take $m_{\rm ex}$=0.7${m}$ with $m_{\rm ex}$ as the free electron mass,  and $m_{\rm ph}$ is the photon mass. The Hopfield coefficients $C$, $X$, and $Y$ values for zero detuning for photonic and excitonic resonances  are $|X|^2=1/2$, $|C|^2=|Y|^2=1/4$ \cite{Byrnes:14} and we note that $\tau_{\rm dip}$ depends on the photon $\tau_{\rm ph}$, direct exciton $\tau_{\rm DX}$, and indirect exciton  $\tau_{\rm IX}$ 
lifetimes, whose values are provided in Table 1. below. Here, $C$, $X$, $Y$ are the photon, direct exciton, and indirect exciton components of the dipolariton wavefunction, respectively \cite{Byrnes:14}.

For the case we consider here, no voltage is applied across the layer, which carries the holes. Although our dipolaritons possess a net electrical charge of zero, the dipole-like spatial separation of the direct exciton and indirect exciton enables us to drive this system with an external electric  field.
In this case effective drive force acting on the dipolaritons is \cite{Serafin17}

\begin{equation}
\bm{F}= e |Y|^2 {E}, 
\end{equation}

The electric field E at $\theta_{E}$ = \ang{0}  is taken as a constant and is directed rightward and parallel to the middle branch of the channel as seen in Fig.1.
We take the dipolariton source in the form of a Gaussian function 
centered  at the base of the stem (see Fig.\ \ref{fig:pattern}) with the full width at half maximum
(FWHM) of 16.7 $\mu$m.

The dipolariton diffusivity is calculated as \cite{Kolmakov:33}
 \begin{equation}
D= \frac{m_{\rm ex}}{m}{|X^{-4}|}D_{\rm ex},
 \end{equation}
 where $D_{\rm ex}$  is the exciton diffusion coefficient. The dipolariton mobility is calculated as \cite{Kolmakov:33}
 \begin{equation}
 \eta _{\rm dip} = \frac{\tau_{\rm dip}}{m},
 \end{equation}
for $\tau_{\rm dip}$ where $\tau_{\rm dip}$ is the momentum relaxation time of the dipolaritons.

To characterize the propagation of the dipolaritons in the channel, we numerically counted the total number of particles $n$, present in each branch of the $\Psi$-shaped channel at locations $x_{\rm n} \geq  450\mu$m, where $x_{\rm n}$  is chosen in such a manner as to only count particles that have propagated at a sufficient distance through the junction of the channel after the branches fork out in Fig.1.

 To characterize the redistribution of the dipolaritons in each branch of the channel, we calculated 
 the fraction of dipolaritons propagating through the upper branch of the channel, or what we define as the efficiency in the channel,

 \begin{equation}
 \varepsilon = {{n_{\rm up} \over{{n_{\rm up} + {n_{\rm mid}+ n_{\rm low}}}}}} \times100 \%.
 \end{equation}

\begin{table}[tb]
\setlength{\arrayrulewidth}{0.2mm}
\setlength{\tabcolsep}{3pt}
\renewcommand{\arraystretch}{1.1}
\centering
\caption{Simulation parameters for a cavity with embedded MoSe$_2$-WS$_2$  bilayer}
\begin{tabular}{|c|c|c|} 
\hline 
\underline{Quantity name} & \underline{Value} & \underline{Variable} \\

Exciton mass & $m_{\rm ex}$ & $0.70 m_{\rm 0}$ \\
Photon mass & $m_{\rm ph}$ & $1.1234\times 10^{-5} m_{\rm 0}$ \\
Dipolariton mass & $m$ & $2.4 \times 10^{-5} m_{\rm 0}$ \\
Dipolariton lifetime  & $\tau_{\rm dip} $ & $15.64 \times 10^{-12}$ s \vspace{-0.1cm}\\
\hspace{0.15cm}  & & \\
Indirect exciton lifetime  & $\tau_{\rm IX} $ & $80 \times 10^{-12}$ s \vspace{-0.1cm}\\
\hspace{0.15cm}  & & \\
Direct exciton lifetime  & $\tau_{\rm DX}$ & $4.0 \times 10^{-12}$ s \vspace{-0.1cm}\\
\hspace{0.15cm}  & & \\
Cavity Photon lifetime  & $\tau_{\rm ph} $ & $100\times 10^{-12}$ s \vspace{-0.1cm}\\
\hspace{0.15cm}  & & \\
Exciton diffusion   & $D_{\rm ex}$ &  $14$ cm$^2$/s \vspace{-0.1cm}\\
\hspace{0.15cm} coefficient & & \\
Dimensionless dipolariton  & $D\times dt/dx^{2}$  &  $271.7$  \vspace{-0.1cm}\\
\hspace{0.15cm} diffusion coefficient & & \\

Dimensionless dipolariton & $\eta_{\rm dip} \times eVdt/dx$ &  $0.0015$  \vspace{-0.1cm}\\
 \hspace{0.15cm} mobility   & & \\
Exciton Energy & $E_{\rm ex}$ & 1.58eV \\
Confining potential  & $\Phi$ & $25-500 meV$\\
Numerical unit of  & $dx$ & 0.15 $\mu$m \vspace{-0.1cm}\\
\hspace{0.15cm}length & & \\
Numerical time step & $dt$ & 9.63 fs \\
\hline
\end{tabular}
\label{tab:params}
\end{table}

where ${n_{\rm up}}$ is the number of dipolaritons in the upper branch of the channel, ${n_{\rm mid}}$ is the number of dipolaritons in the middle branch of the channel, ${n_{\rm low}}$ is the number of dipolaritons in the lower branch of the channel. Fig.2. provides for an illustration of the density of dipolariton particles in the $\Psi$-shaped channel for the specified cases.
 
 We define $ \varepsilon$ in such a manner in order to find the percentage of dipolaritons distributed through the upper branch of the channel relative to the total number of  dipolaritons in the $\Psi$-shaped channel. This enables us to quantify to what extent we can re-route the total number dipolaritons in the channel through the upper branch of the $\Psi$-shaped channel. Thus, we can claim that a higher value for $\varepsilon\ $will indicate a greater ability to drive the dipolaritons through the desired branch of the $\Psi$-shaped channel.

Furthermore, we also investigate the efficiency of the channel when the middle branch of the  $\Psi$-shaped channel is taken to be a buffer, such that the density of dipolaritons  through the middle branch is not considered in the calculation in what we define as the buffered efficiency, $\varepsilon_{2}$. This consideration gives us an efficiency comparison of a $\Psi$-shaped channel to that of Y-shaped channel for the case when the angle between the upper and lower branches is sixty, where for certain cases dipolaritons were able to be re-routed through the upper branch of the channel with greater than 90$\%$ efficiency \cite{Serafin17}. Thus, the buffered efficiency $\varepsilon_{2}$ for the channel in this case is calculated as,

  \begin{equation}
 \varepsilon_{2} = {{n_{\rm up} \over{{n_{\rm up} + {n_{\rm low}}}}}} \times100 \%.
 \end{equation}

\begin{figure}[t]
\hspace{0.5cm}\includegraphics[width=10cm, height=9cm]{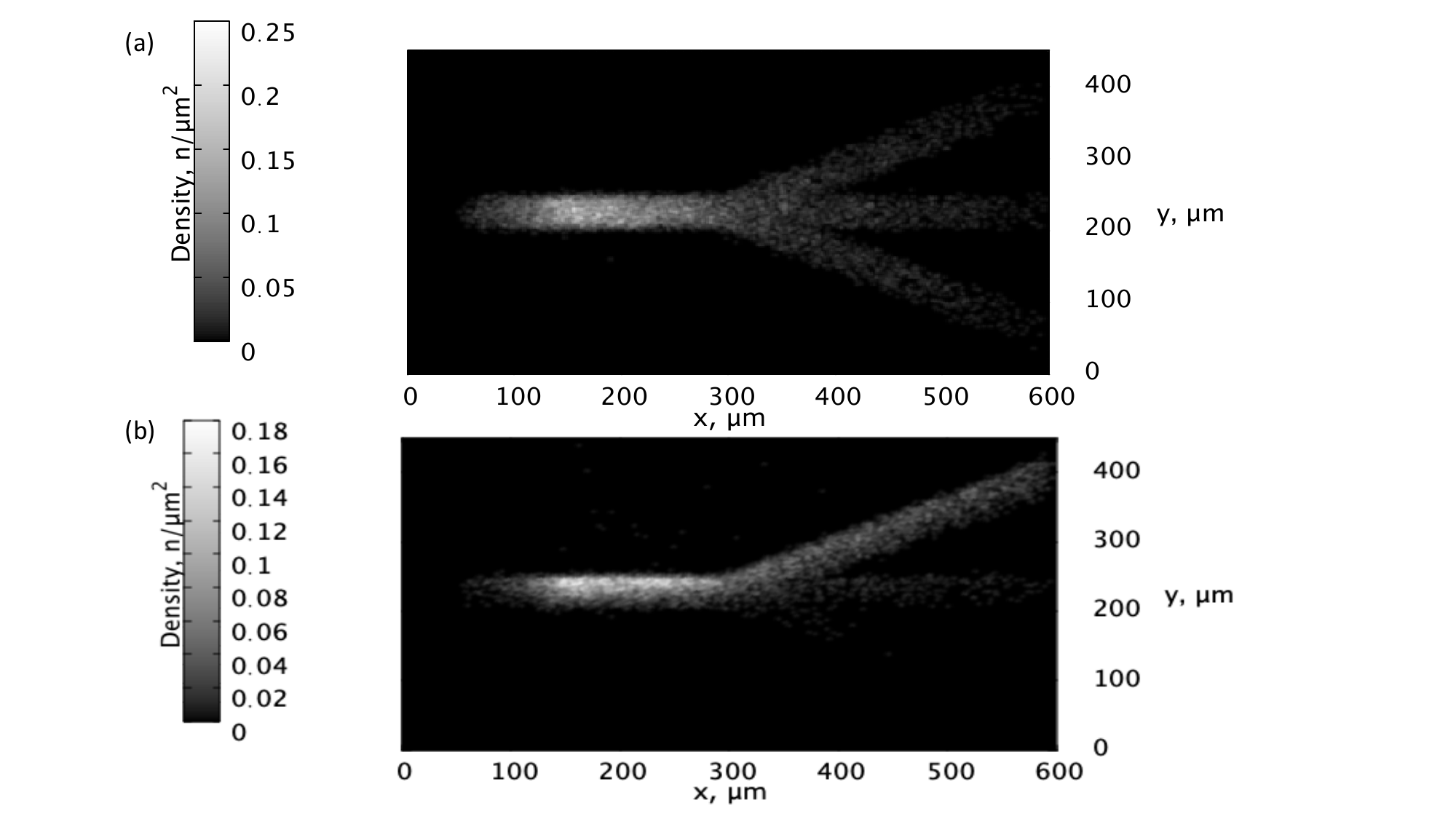}
\caption{Dipolariton propgation through the three branches of the $\Psi\ $channel for the case of (a) $F$=0.5eV/mm and $\theta_{E}$ = 0 (b) $F$=2.0eV/mm and $\theta_{E}$ = 60.}
\label{fig:gaaso120}
\end{figure}

{\section{Optimizing the efficiency of the $\Psi\ $ channel to effectively reroute dipolaritons}}

In order to determine the optimal conditions for directing dipolaritons in the $\Psi$-shaped channel, we varied the electric field angle $\theta_{E}$ in our system and calculated the resulting efficiency in the channel. In order to test the efficacy of changing the electric field angle $\theta_{E}$ on $\varepsilon$, we varied the electric field angle from $\theta_{E}$ = $\ang{0}$ to $\theta_{E}$ = $\ang{60}$. The result of increasing the electric field angle on efficiency at constant driving force can be shown in Fig.3(a)  where we can observe the dipolariton particle number and the channel efficiency as functions of channel electric field angle for the case of a fixed electric driving force of $F$=2.0eV/mm. We see in Fig.3(a)  that an increase in the electric field leads to a monotonic increase in efficiency. In particular we can see that as the electric field angle $\theta_{E}$ is increased from $\theta_{E}$ = \ang{0} to $\theta_{E}$ = \ang{60}, the efficiency $\varepsilon\ $is improved by $\approx$70$\%$. Furthermore, we can observe that total number of particles in the upper branch $n_{up}$ increases from \num{4e+3}  particles to \num{7.1e+3} particles as the electric field angle $\theta_{E}$  is increased from  $\theta_{E}$ = \ang{0} to $\theta_{E}$ = \ang{60}. We can also observe that the number of particles present in the lower branch of the channel $n_{low}$ decreases from \num{3.5e+3} particles to \num{1.3e+2} particles as $\theta_{E}$  is increased from  $\theta_{E}$ = \ang{0} to $\theta_{E}$ = \ang{60}. In addition Fig.3(a) shows that the number of particles going through the middle branch of the channel decreases from \num{4.0e+3} particles to \num{1.3e+3} particles as the electric field angle is increased from $\theta_{E}$ = \ang{0} to $\theta_{E}$ = \ang{60}. Thus, increasing the electric field angle in our $\Psi$-shaped channel from  $\theta_{E}$ = \ang{0} to $\theta_{E}$ = \ang{60} with the driving force $F$ = 2.0eV/mm provides for an efficiency of $\approx$88$\%$. This improvement of efficiency as the electric field angle is increased can be attributed to the fact that as the angle is placed more level to the direction of the upper channel branch the dipolaritons have a greater statistical probability to be driven through the upper and middle branch of the channel rather than the lower branch of the channel. When the electric field angle is set to zero, there is driving towards the middle branch and thus the dipolaritons will propagate primarily towards the middle branch and some will stochastically diffuse into the upper and lower branch at lower populations that the middle branch.  Thus, we can claim that increasing the electric field angle in the channel improves efficiency, while increasing the upper branch population and decreasing the lower branch population of dipolaritons.

\begin{figure}[t]
\hspace{0.5cm}\includegraphics[width=8cm, height=13cm]{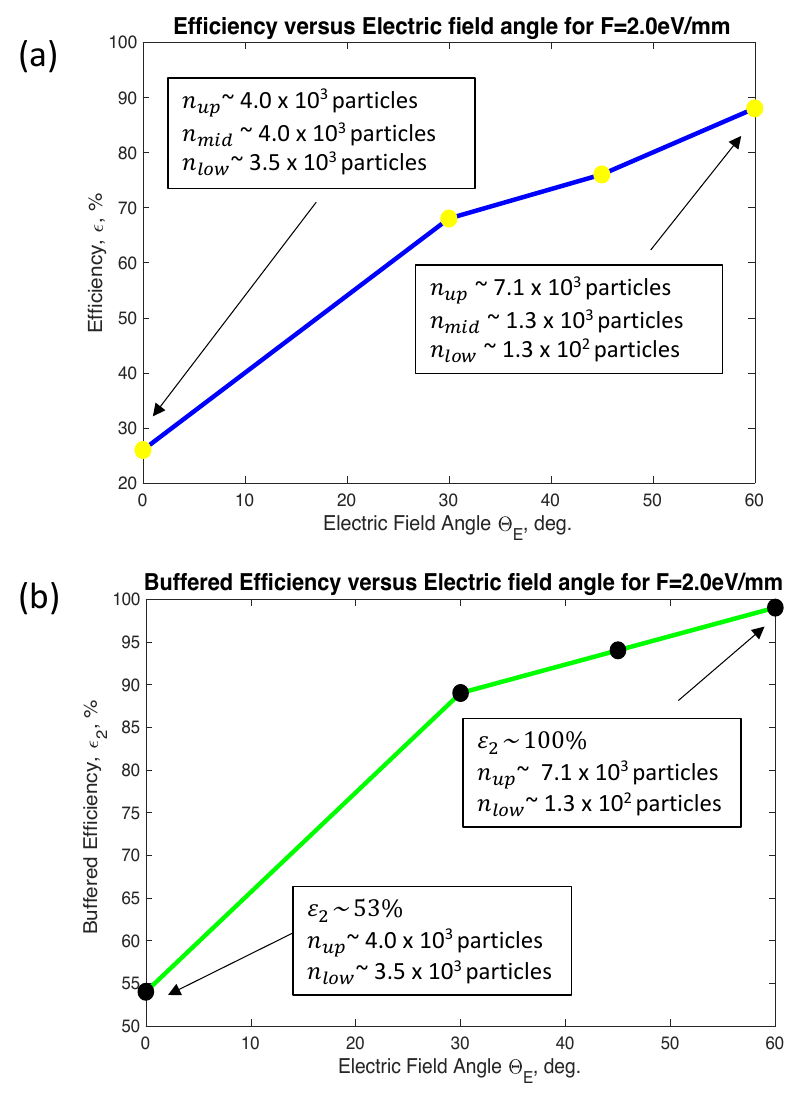}
\caption{ (a) Channel efficiency $\varepsilon\ $ versus the electric field angle $\theta_{E}$ for the case of fixed driving force F = 2.0eV/mm and (b) Buffered channel efficiency $\varepsilon_ {2}\ $versus $\theta_{E}$ for the case of a fixed driving force F = 2.0eV/mm.}
\label{fig:gaaso120}
\end{figure}

\vspace{0.8cm}

In order to determine the efficiency of the channel in the presence of a buffer channel, we numerically calculated the buffered efficiency  $\varepsilon_{2}$ as a function of the electric field angle $\theta_{E}$. In particular, we considered the middle branch of the $\Psi\ $-shaped channel to not be considered in our calculation for $\varepsilon_{2}$ as seen in Eq.(3). In Fig.3(b) we can observe the dipolariton particle number and the channel efficiency as functions of channel electric field angle for the case of a fixed electric driving force of $F$=2.0eV/mm. In particular we can observe that as the electric field angle $\theta_{E}$ is increased from $\theta_{E}$ = \ang{0} to $\theta_{E}$ = \ang{60} the efficiency with buffer $\varepsilon_{2}$ increases monotonically with an increase with the electric field angle $\theta_{E}$. In Fig.3(b) we can see that $\varepsilon_{2}$ is maximized for the case of $\theta_{E}$ = \ang{60} where we can obtain a buffered efficiency of $\varepsilon_{2}$ $\approx$100$\%$ and total particle number of \num{7.1e+3}particles. Thus, increasing the electric field angle from $\theta_{E}$ = \ang{0} to $\theta_{E}$ = \ang{60} provides for  $\approx$ 47$\%$ improvement of $\varepsilon_{2}$ with \num{6.6e+4} particles in the upper branch $n_{up} $. Moreover, we can see that for the case of $\theta_{E}$ = 0  the particles are evenly distributed between the lower and upper branches within the error bars showing fluctuations of particle density in Fig.5. as the channel as geometry would suggest. Compared to the case of no buffer, we see that increasing the electric field angle with a buffer in the $\Psi\ $-shaped channel has a lower impact on the improvement of efficiency, while clearly leaving particle number unchanged. Thus, we can state that considering a buffer in the middle branch of the $\Psi$-shaped channel makes the channel similar geometrically that of a Y-shaped channel with a similar redistribution efficiency \cite{Serafin17}.

\begin{figure}[t]
\hspace{0.5cm}\includegraphics[width=8cm, height=13cm]{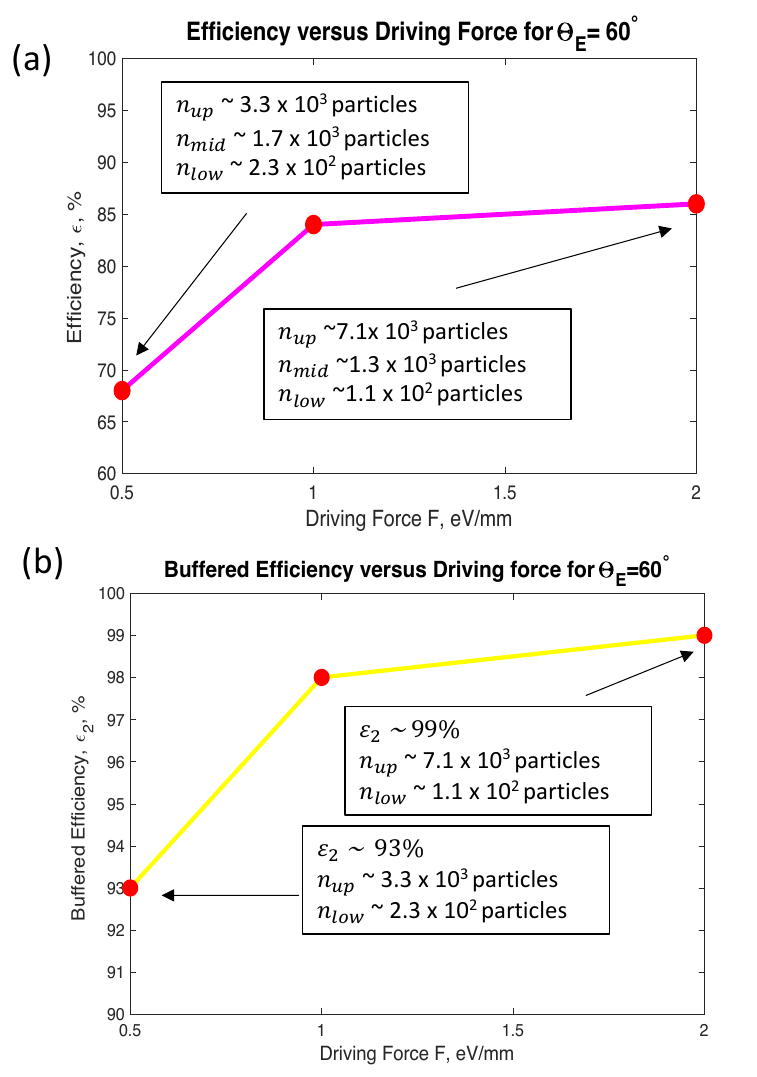}
\caption{(a) Channel efficiency $\varepsilon\ $versus the driving force F for the case of a fixed electric field angle $\theta_{E}$ = \ang{60} and (b) Buffered channel efficiency $\varepsilon_ {2}\ $  versus the driving force F for the case of a fixed electric field angle $\theta_{E}$ = \ang{60}. }
\end{figure}

To determine the optimal conditions for directing dipolaritons in the $\Psi$-shaped channel, we varied the driving force $F$ in our system and studied the resulting distribution of dipolaritons in our system. In Fig.4(a) we can see the efficiency $\varepsilon$ as a function of the driving force $F$ for the case of a fixed electric field angle $\theta_{E}$ = $\ang{60}$. We can observe that the efficiency $\varepsilon$ is a monotonically increasing function with the driving force $F$ where $\varepsilon$ is maximized at a value of $\approx$87$\%$ when the driving force is set to $F$ = 2.0eV/mm. In particular we can see that as the driving force $F$ is increased from 0.5eV/mm to 2.0eV/mm, the efficiency $\varepsilon$ is improved by $\approx$20$\%$. This improvement in efficiency is noted as less substantial than the improvement we can gain by increasing the electric field angle as previously discussed. In addition, we can see in Fig.4(a) that as the driving force $F$ is increased from 0.5eV/mm to 2.0eV/mm, the the number of particles present in the upper branch of the channel increases from \num{3.3e+3} particles to \num{7.1e+3}particles, whilst the number of particles present in the lower branch of the channel decreases from \num{2.3e+2} particles to \num{1.1e+2} particles. This effect can be attributed to the fact that as the electric driving force on the dipolaritons is increased, there are more particles directed towards the direction of the field angle. Thus, we can claim that an increase in the electric field angle in the channel increases the efficiency $\varepsilon$ in the channel, while increasing the total upper branch particle population and lowering the lower branch dipolariton population

In order to determine the efficiency of the channel in the presence of a buffer channel, we numerically calculated the buffered efficiency $\varepsilon_{2}$ as a function of the driving force $F$. In particular, we considered the middle branch of the $\Psi$-shaped channel to not be considered in our calculation for $\varepsilon_{2}$  as seen in Eq.(3). In Fig.4(b) we can see the efficiency $\varepsilon_{2}$ of the $\Psi\ $-shaped channel with a buffer  as a function of the driving force, $F$ for the case of $\theta_{E}$ = \ang{60}. We can observe in Fig.4(b) that the efficiency $\varepsilon_{2}$ is maximized when the driving force $F$ is to set to F = 2.0eV/mm where we find a buffered efficiency of $\approx$99$\%$. Thus we can claim an improvement in buffered efficiency of $\approx$6$\%$ as the driving force $F$ is increased from 0.5eV/mm to 2.0eV/mm. This improvement in efficiency is noticeably less significant when compared to the improvement we obtain with an increase of driving force for the case with no buffer as the starting efficiency for the case of the buffer is already quite high at $\varepsilon_{2}$ $\approx$93$\%$.  Thus, we can report that an increase of the channel electric driving force improves the buffered efficiency; however, this effect is not a substantial as the effect of increasing driving force in the unbuffered channel.

Finally, we seek to further analyze the redistribution of dipolaritons in the case of a buffered channel by defining the performance of the channel P as 

 \begin{equation}
P = {{|n_{\rm up} - n_{\rm low}|\over{{n_{\rm up} + {n_{\rm low}}}}}} \times100 \%.
 \end{equation}

The performance gives us another metric for evaluating how well we can re-route dipolaritons in the case of a buffered channel. The performances for the case of a buffered channel is summarized in Fig. 5. Here the fitting functions for the performances were taken as $\rm{P} = \rm{P}_{max}tanh(\frac{\theta_{E}}{B})$

\begin{figure}[t]

\hspace{0.5cm}\includegraphics[width=7cm, height=7cm]{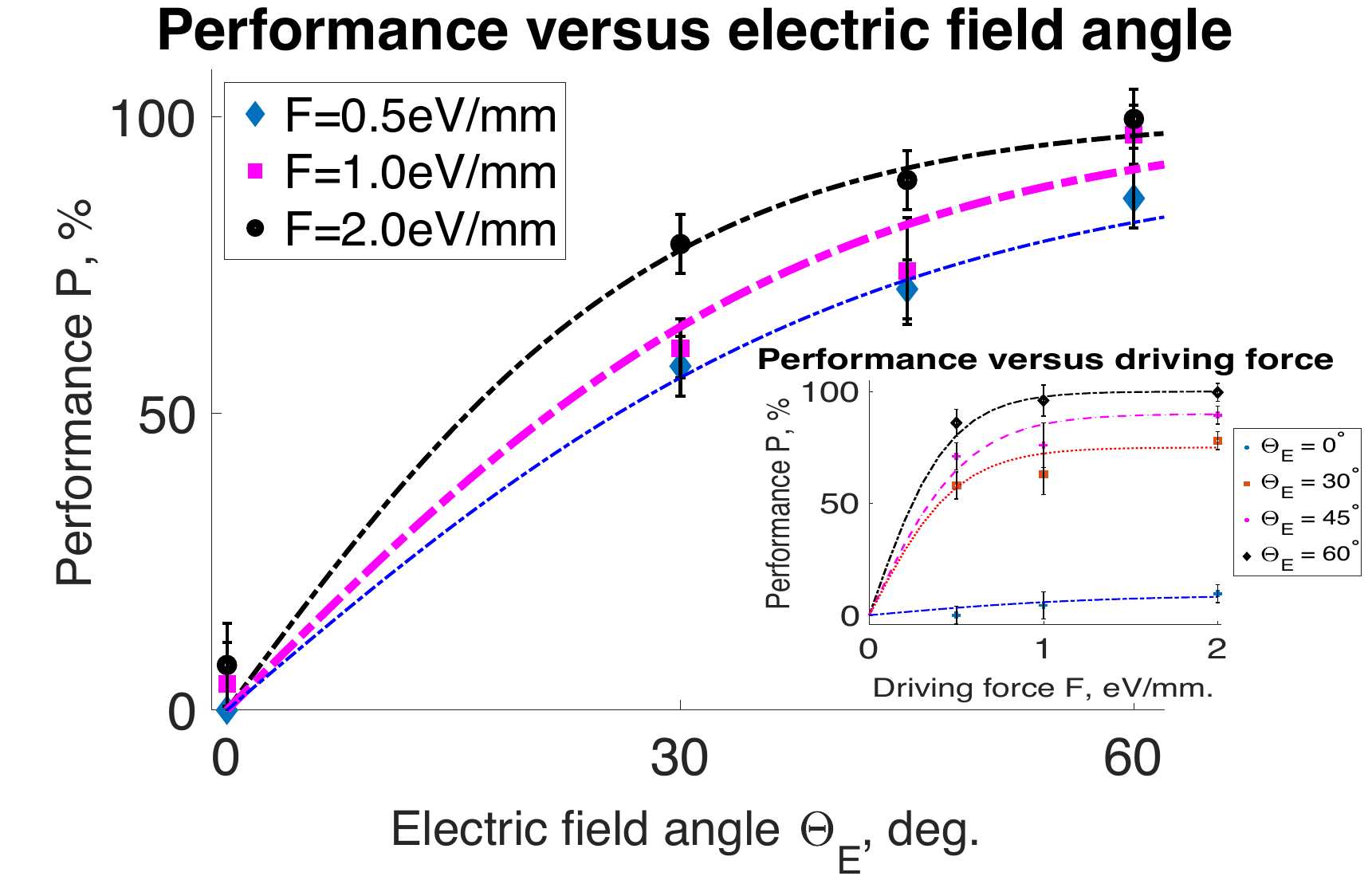}
\caption{Performance in the buffered channel as a function of electric field angle and driving force (inset graph). }

\end{figure}

where $\rm{P}_{max}$ and B are fitting parameters,

The error bars due to fluctuations of particle density were computed from the deviations of our values from the selected fitting parameters, for which we show the optimal performance parameters in Fig.6. We can see the general trend that increasing the electric field angle and driving force in the system provides for monotonically increasing performance P. In particular, for an electric field angle of $\theta_{E}$ = \ang{0}, we obtain a performance of 0 \% within error margins as suggested by geometry. For electric field angles of $\theta_{E}$ = \ang{60}, we obtain performances ranging from 80\%- 99\% within error margins, with an optimal performance attained at $\theta_{E}$ = \ang{60}, $F$= 2.0eV/mm. as also seen in Fig.6.

\vspace{0.8cm}

\begin{figure}[t]
\hspace{0.5cm}\includegraphics[width=8cm, height=7cm]{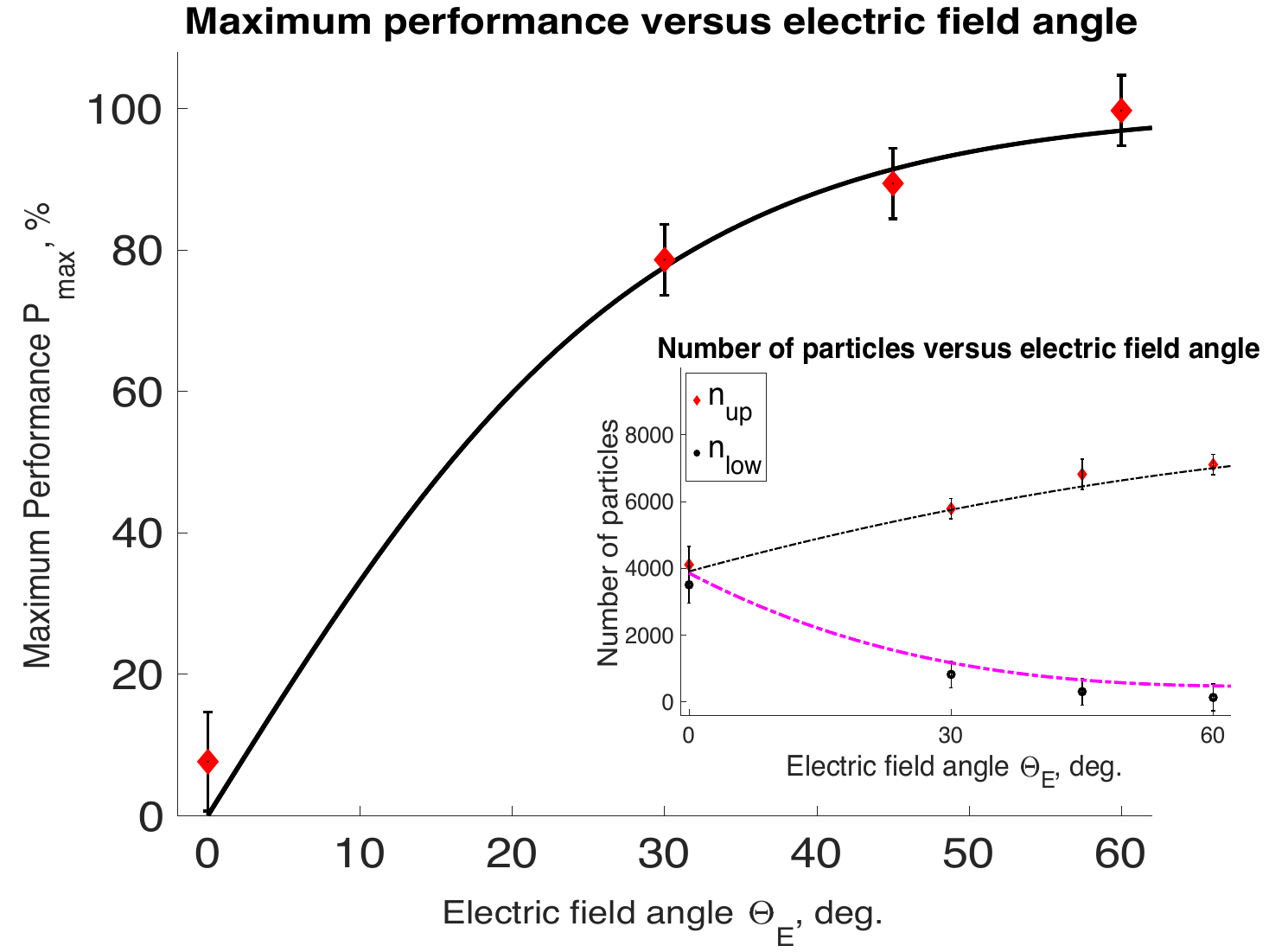}
\caption{Maximum performance versus electric field angle and number of particles versus electric field angle, where the fitting curves are meant as a visual guide for the relationship between particle number and electric field angle (inset graph).}
\end{figure}

\section{Conclusions}
The optimum efficiency $\varepsilon$ for an unbuffered $\Psi$-shaped TMDC channel has been shown to be maximized at particular values of $ \theta_{E}$  and $F$. In particular, for values of $F$ = 2.0eV/mm and $ \theta_{E}$ = \ang{60}, we can re-route $\approx$85$\%$ of dipolaritons through the upper branch of the unbuffered channel. When considering a buffered channel, we are able to re-route $\approx$100$\%$ of dipolaritons as demonstrated in other geometrically similar channels when the channel parameters are set to $F$ = 2.0eV/mm and $ \theta_{E}$ = \ang{60}. For all cases studied, $\varepsilon$ and $\varepsilon_{2}$ were shown to be monotonically increasing functions with an increase of $F$ as well as an increase of $ \theta_{E}$. The impact of increasing driving force and electric field angle was less substantial on $\varepsilon_ {2}$  when compared to $\varepsilon$. These results closely resemble the results when considering Y-shaped channel guides \cite{Serafin17}, with efficiencies and performance over $95\%$ at optimal parameter ranges. However, it is important to note that as our angles approach $ \theta_{E}$ = \ang{60}, we approach a saturation point where a further increase in electric field angle does not generate a substantial increase in upper branch polariton number, thus indicating that a further increase in electric field angle will not necessarily improve performance further. Thus, we can claim that our studies demonstrate optimal efficiencies for the angles between $ \theta_{E}$ = \ang{0} to $\theta_{E}$ = \ang{60}, without commenting on behavior outside of this range. 
	Regarding experimental implementations, there have been investigations into two-terminal devices \cite{Rui} and field-effect transistors \cite{Zhang} that can be run at the voltages we have considered here, thus demonstrating a possibility of experimental testing of the model we have considered here. Our results demonstrate the possibility of expanding the repertoire of other proposed designs of optical transistors in TMDC materials, whilst also opening the route toward the design of novel optoelectronic applications using $\Psi$-shaped channel guides and buffered channels.

\section{Acknowledgments}

 This work was supported in part by the Department of Defense under the grant No. W911NF1810433. The authors are grateful to The Center for Theoretical Physics of New York City College of Technology of The City University of New York for providing computational resources. The authors are also grateful to R. Ya.
 Kezerashvili, O. L. Berman, T. Byrnes for fruitful discussions.

\section{References}

\bibliographystyle{unsrt}
\bibliography{arxivpsi.bib}

\end{document}